# Sparse Transfer Learning for Interactive Video Search Reranking

XINMEI TIAN AND DACHENG TAO, Nanyang Technological University

YONG RUI, Microsoft Advanced Technology Center

Visual reranking is effective to improve the performance of the text-based video search. However, existing reranking algorithms can only achieve limited improvement because of the well-known semantic gap between low level visual features and high level semantic concepts. In this paper, we adopt interactive video search reranking to bridge the semantic gap by introducing user's labeling effort. We propose a novel dimension reduction tool, termed sparse transfer learning (STL), to effectively and efficiently encode user's labeling information. STL is particularly designed for interactive video search reranking. Technically, it a) considers the pair-wise discriminative information to maximally separate labeled query relevant samples from labeled query irrelevant ones, b) achieves a sparse representation for the subspace to encodes user's intention by applying the elastic net penalty, and c) propagates user's labeling information from labeled samples to unlabeled samples by using the data distribution knowledge. We conducted extensive experiments on the TRECVID 2005, 2006 and 2007 benchmark datasets and compared STL with popular dimension reduction algorithms. We report superior performance by using the proposed STL based interactive video search reranking.

1. INTRODUCTION

With the development of video recording and storage devices, as well as the improvement of transmission and compression techniques, the number of videos on Web increases explosively. Meanwhile, video-sharing websites become more and more popular. For example, there are hundreds of millions of videos on Youtube, Tudou, and Youku. It reports that Youtube serves more than 1 billion video views every day. As a consequence, efficient and effective video search tools are essential for web surfing. The most-frequently used video search engines, e.g., Google, Microsoft's Bing, Yahoo, retrieve videos by indexing their associated textual information, such as video tags, captions, surrounding texts in Web pages and speech transcripts. However, the performance of the text-based video search is unsatisfactory because textual information cannot describe the video's rich content comprehensively and substantially. As a consequence, techniques with essential visual information involved are proposed to build video search prototypes. One kind of such technique is the content-based annotation [Carneiro et al. 2007; Qi et al. 2008; Jeon et al. 2003; Natsev et al. 2007; Yang 2010] which is used to enrich the textual descriptions for videos. However, the scalability and accuracy of automatic annotation are far from satisfactory for large-scale video datasets. Another kind of technique is content-based video retrieval (CBVR) [Chang et al. 1999; Lew et al. 2006; Nguyen and Worring 2008] in which textual descriptions are not required and, only visual features are used. However, the query examples, which are essentially required by CBVR, are often unavailable for users. Besides, it has been acknowledged that pure content-based video retrieval

Authors' addresses: X. Tian and D. Tao, School of Computer Engineering, Nanyang Technological University, 50 Nanyang Avenue, Blk N4, Singapore 639798; email: xinmeitian@gmail.com , dacheng.tao@gmail.com . Y. Rui, Microsoft China R&D Group, 2F, Sigma Center, Microsoft Advanced Technology Center, 49 Zhichun Road, Haidian District, Beijing, China 100190; email: yongrui@microsoft.com.

This work was funded by the Open Project Program of the State Key Lab of CAD&CG (Grant No. A1006), Zhejiang University.

DOI10.1145/0000000.0000000 http://doi.acm.org/10.1145/0000000.0000000

approaches cannot work well, due to the semantic gap [Smeulders et al. 2000] between low level visual features and high level semantic concepts.

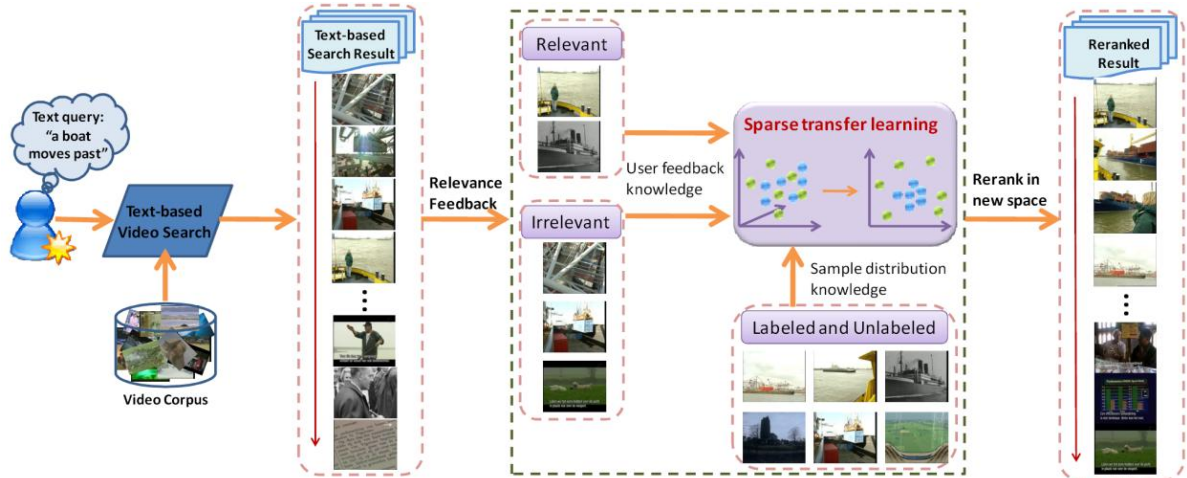

Fig. 1. The framework of the sparse transfer learning based interactive video search reranking.

To take the merits of both textual and visual information for video search, an integrated framework is demanded. Visual reranking [Hsu et al. 2006; Jing and Baluja 2008; Tian et al. 2008] is such a suitable technique which utilizes visual information to refine text-based video search results. Particularly, visual reranking first conducts the text-based search to return a coarse result from large indexing due to efficiency. Afterward, top returned results are reordered by using the visual information. However, due to the semantic gap, the visual reranking methods which use general (query independent) low-level visual feature can only slightly improve the performance. Since relevance feedback has been proven effective to reduce the semantic gap, in this paper, we introduce relevance feedback to video search reranking, termed interactive video search reranking. Figure 1 illustrates the framework. After the text-based search engine returns a coarse result, the user is required to label some samples in the returned set as query relevant or irrelevant. The labeling information indeed reflects user's search intention. We then propose a new dimension reduction tool, termed sparse transfer learning (STL), to effectively and efficiently encode user's search intention. By reranking samples in the subspace learned by STL, we can significantly improve the search performance.

The interactive video search reranking is different from the conventional relevance feedback techniques [Rocchio 1971; Rui et al. 1998; Lew et al. 2006; Tao et al. 2006; Rui et al. 1997] in image and video retrieval. In image and video retrieval, samples are general images/video in several categories and are distributed widely. On the contrary, in interactive video search reranking, the distribution of the samples returned by text-based search engines reflects the information provided by the textual feature cue. Therefore, it is necessary to design a specific tool to encode this special distribution information into visual representation. None of the existing dimension reduction methods is designed specifically for video reranking problem and cannot be well generalized for it. Therefore, a new dimension reduction method, i.e., sparse transfer learning, is developed in this paper to meet the specific requirement in interactive reranking.

Figure 1 shows that STL exploits two pieces of knowledge data, i.e., the *user's feedback knowledge* contained in labeled samples domain and the *sample distribution knowledge* contained in all samples (labeled and unlabeled). The objective subspace is learned by transferring the user feedback knowledge from labeled samples to unlabeled samples by preserving the sample distribution knowledge, i.e., the prior information provided by the text-based search. The number of labeled samples is much less than the visual feature dimension, so it is necessary to control the model complexity according to the regularization theory [Neumaier 1998]. In this paper, we introduce the elastic net penalty [Zou and Hastie 2005] to control the volume of the objective function to obtain a sparse representation of the subspace. Supporting experiments show STL is effective for interactive video search reranking. In summary, we have three major technical contributions in this paper:

1) We propose a transfer learning based dimension reduction tool that propagates user's feedback knowledge (labels) from labeled samples to unlabeled samples by using data distribution knowledge.

2) We define a pair-wise discriminant analysis (PDA) to separate labeled relevant samples from labeled irrelevant ones and preserve the local structure of the data distribution.

3) We introduce the sparse penalty to the proposed transfer leaning based dimension reduction to explain the projection matrix psychologically and physiologically, to save the time and space cost, and to reduce over-fitting.

The remainder of this paper is organized as follows. Section 2 reviews related works on reranking and dimension reduction. We detail the proposed sparse transfer learning for dimension reduction in Section 3. Section 4 discusses the relationship between PDA and biased discriminant analysis (BDA) [Zhou and Huang 2001]. The time and space complexities of STL are analyzed in Section 5. Extensive experiments and analyses are presented in Section 6. Section 7 concludes the paper.

2. RELATED WORK

Visual reranking has been proven to be an effective tool to improve the text-based search result by incorporating visual information. A number of visual reranking methods have been proposed in recent years. These methods can be divided into three categories.

The first category is classification-based [Yan and Hauptmann 2004]. This kind of methods selects the pseudo -positive and pseudo-negative samples from the initial text -based search results and then trains a classifier with these samples. The video shots are reranked according to the predictions given by the trained classifier.

The second category is clustering-based. In [Hsu et al. 2006], it assigned each video shot a soft pseudo label according to the initial text-based search ranking score and then the Information Bottleneck principle [Slonim and Tishby 1999] was applied to finding the optimal clustering which maximizes the mutual information between the clusters and the labels. The reranked list was achieved by ordering the clusters according to the cluster conditional probability firstly and then ordering the samples within a cluster based on their local feature density estimated via kernel density estimation. Liu et al. (2008) first grouped visually similar videos into clusters and then ranked all clusters according to their textual similarity to the query term.

The third category for visual reranking is graph-based. Jing and Baluja (2008) applied the well-known pagerank algorithm to image search reranking by directly treating images as documents and their visual similarities as probabilistic hyperlinks. Hsu et al. (2007) modeled the reranking process as a random walk over a graph that is constructed by using samples (video shots) as the nodes and the edges between them being weighted by visual similarities. The ranking scores are propagated through the edges and the stationary probability of the random walk process is used as the reranked score directly. A multi-graph extension is presented in [Liu et al. 2007] to further the performance.

In all visual reranking methods, the visual features play a key role. Popular visual features in visual reranking include global features, e.g., color moments [Hsu et al. 2006; Tian et al. 2008] and Gabor texture [Hsu et al. 2006], and local features, e.g., scale invariant feature transform (SIFT) [Jing and Baluja 2008]. However, due to the semantic gap, visual reranking can only slightly improve the performance for video search. The visual features describe the video shots query independently and thus cannot reflect the context information of this query. For example, we expect two video shots, one for sheep on a hill and another for cows on a hill, have similar visual descriptions for query "*hills or mountains*" but dissimilar visual descriptions for query "*sheep or goats*". The commonly used visual features fail to capture this query dependent property.

By introducing user's labeling information, we can reduce the semantic gap by effectively learning a query dependent visual representation to encode the contextual information. The dimension reduction, which aims to find a compact representation for the samples in a low dimensional subspace, is a suitable candidate. In the past decades, a dozen of dimension reduction algorithms have been proposed, e.g., principal components analysis (PCA) [Hotteling 1933], nonlinear discriminant analysis [Yan et al. 2007], and manifold regularized discriminative non-negative matrix factorization [Guan et al. 2011]. However, it is problematic to directly use existing dimension reduction methods to learn query dependent visual features for video search reranking. Unsupervised dimension reduction methods, e.g., PCA, locally linear embedding (LLE) [Roweis and Saul 2000], exploit a subspace or submanifold on the whole sample space with the user's labeling information ignored. As a consequence, these algorithms fail to capture the contextual information of the query. Supervised algorithms, e.g., Fisher's linear discriminant analysis (LDA) [Fisher 1936], BDA [Zhou and Huang 2001], supervised locality preservation projections (SLPP) [Cai et al. 2005] and nonlinear discriminant analysis [Yan et al. 2007], learn a subspace on the labeled set so they ignore the structural information of all samples. Besides, usually only a few labeled samples are available in interactive video search reranking and thus the subspace learned with supervised dimension reduction methods may bias to the space spanned by the labeled samples. Therefore some semi-supervised algorithms, e.g., semi-supervised discriminant analysis (SDA) [Cai et al. 2007] and semantic manifold learning (SML) [Lin et al. 2005], have been developed to model both labeled and unlabeled samples. S. Si et al (2010) introduced a Bregman divergence-based regularization term to minimize the distribution difference between training and testing samples. Tian et al. (2010) proposed a local-global discriminant dimension reduction for web image search reranking. Recently, sparse subspace learning draws increasing interests and many dimension reduction methods are extended to their sparse version. For example, Hou et al. (2004) proposed sparse PCA. Cai et al. (2007) proposed a general framework for sparse projections learning. Within this framework, several dimension reduction methods, e.g. LDA [Fisher 1936], LPP [He and Niyogi 2003], derive their sparse solution.

However, these methods are not suitable for interactive video search reranking since they are not particularly designed for video search reranking problem. In this paper, we proposed a new method, named sparse transfer learning (STL). The specificity of STL for interactive video search reranking lies in 3-fold. First, STL maintains the dominant structure of sample distribution. This prior distribution information is very useful in video reranking because they reflect the information derived from the textual feature cue. Second, STL encodes user feedback knowledge by utilizing a pair-wise discriminant analysis to separate labeled relevant samples from labeled irrelevant ones. Third, considering the insufficiency of labeled samples (especially the labeled relevant samples), STL introduces elastic net to avoid the over-fitting problem and to provide a good interpretation of the subspace projection.

3. SPARSE TRANSFER LEARNING

In visual reranking, for a query the top-N video shots $I = \{I_1, \cdots, I_N\}$ returned by the text-based search engine are considered for further processing. We represent their visual information by using low level features, i.e., $\mathbf{X} = [\mathbf{x}_1, \cdots, \mathbf{x}_N] \in \mathbf{R}^{m \times N}$ with an m-dimensional visual feature vector $\mathbf{x}_i \in \mathbf{R}^m$ for $I_i$. The performance of reranking directly based on feature $\mathbf{X}$ is usually poor because of the gap between the low level visual features and high level semantics. With user's interactions, this semantic gap can be reduced significantly. By mining user's labeling information, we can learn a submanifold to encode the user's intention. This submanifold is embedded in the ambient space, i.e., the high dimensional visual feature space $\mathbf{R}^m$. In this paper, a linear subspace $\mathbf{U} = [\mathbf{u}_1, \cdots, \mathbf{u}_d] \in R^{m \times d}$ is used to approximate this submanifold and then samples can be represented as $\mathbf{Y} = \mathbf{U}^T \mathbf{X} = [\mathbf{y}_1, \cdots, \mathbf{y}_N] \in \mathbf{R}^{d \times N}$ (d<m), wherein $\mathbf{y}_i \in \mathbf{R}^d$ is the low dimensional representation of $I_i$. By using $\mathbf{Y}$, an improved reranking result can be obtained.

3.1 User Feedback Knowledge

To learning the query dependent visual representation, we first need to know which video shots are query relevant and which are irrelevant. Therefore, relevance feedback is first conducted to collect necessary information to localize the user-driven semantic space. With the labeled samples, we exploit the discriminative information to grasp the contextual information.

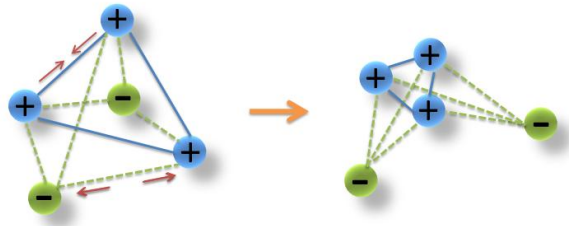

Fig. 2. In PDA, we require the relevant-relevant sample pairs to be as close as possible and the relevant-irrelevant sample pairs to be as far away as possible in the learned subspace.

Many discriminative analysis methods have been proposed, such as, LDA [Fisher 1936], BDA [Zhou and Huang 2001], MMDA [Bian and D. Tao 2011]. Here, by investigating the distribution of the video shots returned by the text-based search engine, it is often observed that the relevant video shots are often similar to each other while the irrelevant ones are irrelevant in its own way. Therefore, we propose a pair-wise discriminative analysis (PDA) method to learn discriminative information, as illustrated in Fig. 2. To maximally separate the relevant samples from irrelevant ones and learn a shared subspace occupied by the relevant ones, we require the relevant-relevant sample pairs to be as close as possible and the relevant-irrelevant sample pairs to be as far away as possible in the learned subspace,

$$\min\left(\frac{1}{N_+N_+}\sum_{I_i\in+}\sum_{I_j\in+}\|\mathbf{y}_i-\mathbf{y}_j\|^2 - \beta\frac{1}{N_+N_-}\sum_{I_i\in+}\sum_{I_k\in-}\|\mathbf{y}_i-\mathbf{y}_k\|^2\right), \qquad (1)$$

where sets "$+$" and "$-$" denote the labeled relevant and labeled irrelevant samples. The $N_+$ and $N_-$ are the number of the samples in these two sets respectively. The $\beta$ is a trade-off parameter to control the influence of the two parts.

We rewrite the objective function in (1) in another form for better understanding,

$$\begin{aligned}
&\frac{1}{N_+N_+}\left(\sum_{I_i\in+}\sum_{I_j\in+}\|\mathbf{y}_i-\mathbf{y}_j\|^2 - \beta\frac{N_+}{N_-}\sum_{I_i\in+}\sum_{I_k\in-}\|\mathbf{y}_i-\mathbf{y}_k\|^2\right)\\
&=\frac{1}{N_+N_+}\sum_{I_i\in+}\left(\sum_{I_j\in+}\|\mathbf{y}_i-\mathbf{y}_j\|^2 - \beta'\sum_{I_k\in-}\|\mathbf{y}_i-\mathbf{y}_k\|^2\right)\\
&=\frac{1}{N_+N_+}\sum_{I_i\in+}t_i,
\end{aligned} \qquad (2)$$

where $\beta'=\beta\frac{N_+}{N_-}$ and $t_i=\sum_{I_j\in+}\|\mathbf{y}_i-\mathbf{y}_j\|^2 - \beta'\sum_{I_k\in-}\|\mathbf{y}_i-\mathbf{y}_k\|^2$. Minimizing $t_i$ means that, for each relevant sample $I_i\in+$, we expect it to be close to all the other relevant ones but to be far away from all irrelevant ones.

Our previous work, the patch alignment framework [Zhang et al. 2008], unifies popular dimension reduction algorithms, e.g., LDA [Fisher 1936], LLE [Roweis and Saul 2000], ISOMAP [Tenenbaum et al. 2000], and locality preserving projections (LPP) [He and Niyogi 2003], into a general framework. The proposed PDA is also developed under this framework. It contains two stages: part optimization and whole alignment.

We build a local patch for each positive sample and derive the corresponding $L_i$ from $t_i$. Define a coefficient vector $w_i = [\underbrace{1,\cdots,1}_{N_+},\underbrace{-\beta',\cdots,-\beta'}_{N_-}]$ and the corresponding low dimensional feature matrix is given by $\mathbf{Y}_i=[\mathbf{y}_i,\mathbf{y}_1^+,\cdots\mathbf{y}_{N_+}^+,\mathbf{y}_1^-,\cdots\mathbf{y}_{N_-}^-]$, wherein each $\mathbf{y}_i^+$ is the $i$th sample in relevant set "$+$" and each $\mathbf{y}_i^-$ is the $i$th sample in irrelevant set "$-$". Then, we can rewrite $t_i$ as

$$\begin{aligned}
t_i &= \sum_{I_j\in+}\|\mathbf{y}_i-\mathbf{y}_j\|^2 - \beta'\sum_{I_k\in-}\|\mathbf{y}_i-\mathbf{y}_k\|^2\\
&= \sum_{k=1}^{N_++N_-}(w_i)_k\|(\mathbf{Y}_i)_1-(\mathbf{Y}_i)_{k+1}\|^2\\
&= \mathrm{tr}\left(\mathbf{Y}_i\begin{bmatrix}-e_{N_L}^T\\ I_{N_L}\end{bmatrix}\mathrm{diag}(w_i)\begin{bmatrix}-e_{N_L} & I_{N_L}\end{bmatrix}\mathbf{Y}_i^T\right)\\
&= \mathrm{tr}(\mathbf{Y}_i\mathbf{L}_i\mathbf{Y}_i^T),
\end{aligned} \qquad (3)$$

where $(\mathbf{Y}_i)_j$ is the $j$th column of $\mathbf{Y}_i$, $N_L=N_++N_-$, $\mathbf{L}_i=\begin{bmatrix}-e_{N_L}^T\\ I_{N_L}\end{bmatrix}\mathrm{diag}(w_i)\begin{bmatrix}-e_{N_L} & I_{N_L}\end{bmatrix}$ with $e_{N_L}=[1,\cdots,1]^T\in R^{N_L}$.

Each patch $\mathbf{Y}_i$ has its own coordinate system and all $\mathbf{Y}_i$s can be aligned together into a consistent coordinate via selection matrices. Assuming the coordinate of patch $\mathbf{Y}_i$ is selected from the global coordinate $\mathbf{Y} = [\mathbf{y}_1, \mathbf{y}_2, \cdots, \mathbf{y}_N] \in R^{d \times N}$, $\mathbf{Y}_i$ can be rewritten as $\mathbf{Y}_i = \mathbf{Y}\mathbf{S}_i$. The selection matrix $\mathbf{S}_i \in R^{N \times (k+1)}$ is defined as [Zhang and Zha 2004]

$$(\mathbf{S}_i)_{pq} = \begin{cases} 1, & \text{if } p = F_i(q) \\ 0, & \text{else} \end{cases}, \qquad (4)$$

where $F_i = [i, i_1, \cdots, i_k]$ is the index vector for samples in $\mathbf{Y}_i$. More details about patch alignment can refer to [Zhang et al. 2008].

Align all local patches for relevant samples together and then we have

$$\begin{aligned}\sum_{I_i \in +} \mathrm{tr}\left(\mathbf{Y}_i \mathbf{L}_i \mathbf{Y}_i^\mathrm{T}\right) &= \sum_{I_i \in +} \mathrm{tr}\left(\mathbf{Y}\mathbf{S}_i \mathbf{L}_i \mathbf{S}_i^\mathrm{T} \mathbf{Y}^\mathrm{T}\right) \\ &= \mathrm{tr}\left(\mathbf{Y} \sum_{I_i \in +} (\mathbf{S}_i \mathbf{L}_i \mathbf{S}_i^\mathrm{T}) \mathbf{Y}^\mathrm{T}\right) \\ &= \mathrm{tr}\left(\mathbf{Y}\mathbf{L}\mathbf{Y}^\mathrm{T}\right),\end{aligned} \qquad (5)$$

where $\mathbf{L} = \sum_{I_i \in +} (\mathbf{S}_i \mathbf{L}_i \mathbf{S}_i^\mathrm{T})$. Substituting (5) into (2) we have

$$\begin{aligned}&\min \frac{1}{N_+ N_+} \sum_{I_i \in +} \sum_{I_j \in +} \|\mathbf{y}_i - \mathbf{y}_j\|^2 - \beta \frac{1}{N_+ N_-} \sum_{I_i \in +} \sum_{I_k \in -} \|\mathbf{y}_i - \mathbf{y}_k\|^2 \\ &= \min \frac{1}{N_+ N_+} \mathrm{tr}\left(\mathbf{Y}\mathbf{L}\mathbf{Y}^\mathrm{T}\right).\end{aligned}$$

### 3.2 Sample Distribution Knowledge

In most cases, only a few labeled samples can be obtained because the users will lose their patience for labeling very soon. As a consequence, the subspace learned with only discriminative information may bias to that spanned by labeled samples. Beyond labeled samples, there are a large number of unlabeled samples that can be used to correct this bias. The prior information of the sample distribution is very useful in reranking because they reflect the information derived from the textual feature cue. It is widely assumed that more often appeared samples have higher probability to be relevant. The clustering -based [Hsu et al. 2006] and graph-based reranking methods [Hsu et al. 2007; Jing and Baluja 2008; Liu et al. 2007] are all developed based on this assumption. To incorporate the sample distribution information into the feature learning process, we maintain the dominant structure of the distribution after dimension reduction.

PCA, which keeps the subspace that has largest variance, i.e., the principle components, is a suitable candidate. Therefore, we minimize the distance between the objective subspace and that obtained by PCA while preserve the discriminative information from labeled samples. PCA is conducted to obtain a low dimensional representation $\mathbf{M}$, and then we try to maintain the dominant structure of the whole dataset by restricting the objective feature $\mathbf{Y}$ to be similar to $\mathbf{M}$, i.e.,

$$\min \frac{1}{N} \|\mathbf{M} - \mathbf{Y}\|^2.$$

To transfer the discriminative information learning from the labeled samples, we connect the labeled samples and the unlabeled ones in the learned subspace by restricting their latent structure to be consistent with that in the ambient feature space. A trade-off parameter $\alpha$ is introduced to combining the two parts,

$$\min \frac{1}{N} \|\mathbf{M} - \mathbf{Y}\|^2 + \alpha \frac{1}{N_+ N_+} \mathrm{tr}\left(\mathbf{Y}\mathbf{L}\mathbf{Y}^\mathrm{T}\right). \qquad (6)$$

By optimizing (6), the discriminative information in the labeled sample domain and the sample distribution information in the whole dataset can be transferred from labeled samples to the unlabeled ones.

### 3.3 Sparsity Penalty

In (6), we aim to learn the subspace wherein each basis is a linear combination of all features in the ambient space, i.e., all (or most of) elements in $\mathbf{U}$ are non-zero. In interactive video search reranking, the number of labeled samples is much less than the visual feature dimension, so it is necessary to control the model complexity according to the regularization theory [Neumaier 1998]. To control the volume of (6), we introduce the elastic net penalty [Zou and Hastie 2005] to the objective function and thus we can obtain a sparse representation of the subspace $\mathbf{U}$,

$$\min \frac{1}{N}\|\mathbf{M}-\mathbf{Y}\|^2 + \alpha \frac{1}{N_+ N_+} \operatorname{tr}\left(\mathbf{YLY}^{\mathrm{T}}\right) + \frac{\lambda_1}{N}\|\mathbf{U}\|_1 + \frac{\lambda_2}{N}\|\mathbf{U}\|^2$$

$$= \min \|\mathbf{M}-\mathbf{Y}\|^2 + \alpha \frac{N}{N_+ N_+} \operatorname{tr}\left(\mathbf{YLY}^{\mathrm{T}}\right) + \lambda_1 \|\mathbf{U}\|_1 + \lambda_2 \|\mathbf{U}\|^2$$

$$= \min \|\mathbf{M}-\mathbf{Y}\|^2 + \alpha' \operatorname{tr}\left(\mathbf{YLY}^{\mathrm{T}}\right) + \lambda_1 \|\mathbf{U}\|_1 + \lambda_2 \|\mathbf{U}\|^2 ,$$

where $\lambda_1$ and $\lambda_2$ are two parameters to control the 1-norm penalty and 2-norm penalty. The learned sparse subspace $\mathbf{U}$ enjoys the following advantages:

1) sparsity can make the subspace $\mathbf{U}$ more succinct and simpler, and thus the subsequent calculation becomes more efficient. Parsimony is especially an important factor when the dimension of the original samples is very high;

2) sparsity can control the weights of original features and decrease the variance brought by possible over-fitting with the least increment of the bias. Therefore, the learned subspace $\mathbf{U}$ can generalize better; and

3) sparsity provides a good interpretation of the subspace $\mathbf{U}$, and thus reveals an explicit relationship between the objective of the model and the given variables. This is important because we can understand the query better by learning which kind of feature plays more important role for different kinds of queries.

### 3.4 Solution for Sparse Transfer Learning

Up to now, we have obtained the overall objective function of the proposed STL for dimension reduction,

$$\min \|\mathbf{M}-\mathbf{Y}\|^2 + \alpha' \operatorname{tr}\left(\mathbf{YLY}^{\mathrm{T}}\right) + \lambda_1 \|\mathbf{U}\|_1 + \lambda_2 \|\mathbf{U}\|^2 . \tag{7}$$

It is not easy to find the optimal sparse solution for problem (7). In this paper, we show that the objective function is of a quadratic form with 1-norm penalty and thus the least angle regression (LARS) [Efron et al. 2004], one of the most popular algorithms in sparse learning, can be utilized to obtain the optimal solution. We substitute $\mathbf{Y} = \mathbf{U}^{\mathrm{T}}\mathbf{X}$ into (7), and get

$$\min \|\mathbf{M}-\mathbf{U}^{\mathrm{T}}\mathbf{X}\|^2 + \alpha' \operatorname{tr}\left(\mathbf{U}^{\mathrm{T}}\mathbf{XLX}^{\mathrm{T}}\mathbf{U}\right) + \lambda_1 \|\mathbf{U}\|_1 + \lambda_2 \|\mathbf{U}\|^2$$

$$= \min \operatorname{tr}\left((\mathbf{M}-\mathbf{U}^{\mathrm{T}}\mathbf{X})(\mathbf{M}-\mathbf{U}^{\mathrm{T}}\mathbf{X})^{\mathrm{T}}\right) + \alpha' \operatorname{tr}\left(\mathbf{U}^{\mathrm{T}}\mathbf{XLX}^{\mathrm{T}}\mathbf{U}\right) + \lambda_1 \|\mathbf{U}\|_1 + \lambda_2 \|\mathbf{U}\|^2$$

$$= \min \operatorname{tr}\left(\mathbf{U}^{\mathrm{T}}\mathbf{X}(\alpha'\mathbf{L}+\mathbf{I})\mathbf{X}^{\mathrm{T}}\mathbf{U} - \mathbf{M}\mathbf{X}^{\mathrm{T}}\mathbf{U} - \mathbf{U}^{T}\mathbf{X}\mathbf{M}^{\mathrm{T}}\right) + \lambda_1 \|\mathbf{U}\|_1 + \lambda_2 \|\mathbf{U}\|^2 .$$

We use $\mathbf{A}$ to denote $(\alpha'\mathbf{L}+\mathbf{I})$ for simplicity and obtain

$$\min \operatorname{tr}\left(\mathbf{U}^{\mathrm{T}}\mathbf{XAX}^{\mathrm{T}}\mathbf{U} - \mathbf{M}\mathbf{X}^{\mathrm{T}}\mathbf{U} - \mathbf{U}^{T}\mathbf{X}\mathbf{M}^{\mathrm{T}}\right) + \lambda_1 \|\mathbf{U}\|_1 + \lambda_2 \|\mathbf{U}\|^2 . \tag{8}$$

The $\mathbf{A}$ is symmetric, so we have $\mathbf{A} = \mathbf{VDV}^{\mathbf{T}}$ where $\mathbf{V}$ and $\mathbf{D}$ are the eigenvector and eigenvalue matrix of the eigendecomposition respectively. Then we can rewrite the first part in (8) as

$$\operatorname{tr}\left(\mathbf{U}^{\mathrm{T}}\mathbf{XAX}^{\mathrm{T}}\mathbf{U} - \mathbf{M}\mathbf{X}^{\mathrm{T}}\mathbf{U} - \mathbf{U}^{T}\mathbf{X}\mathbf{M}^{\mathrm{T}}\right)$$

$$= \operatorname{tr}(\mathbf{U}^{\mathrm{T}}\mathbf{X}(\mathbf{VD}^{1/2})(\mathbf{D}^{1/2}\mathbf{V}^{\mathrm{T}})\mathbf{X}^{\mathrm{T}}\mathbf{U} - \mathbf{M}(\mathbf{VD}^{1/2})(\mathbf{VD}^{1/2})^{-1}\mathbf{X}^{\mathrm{T}}\mathbf{U} - \mathbf{U}^{T}\mathbf{X}(\mathbf{VD}^{1/2})(\mathbf{VD}^{1/2})^{-1}\mathbf{M}^{\mathrm{T}})$$

$$= \|(\mathbf{VD}^{1/2})^{-1}\mathbf{M}^{\mathrm{T}} - (\mathbf{D}^{1/2}\mathbf{V}^{\mathrm{T}})\mathbf{X}^{\mathrm{T}}\mathbf{U}\|^2 - \operatorname{tr}\left(\mathbf{MA}^{-1}\mathbf{M}^{\mathrm{T}}\right)$$

where $\operatorname{tr}\left(\mathbf{MA}^{-1}\mathbf{M}^{\mathrm{T}}\right)$ is a constant item and can be ignored for optimization.

Then we can further rewrite (8) as

$$\min \left\| (\mathbf{V}\mathbf{D}^{1/2})^{-1}\mathbf{M}^{\mathrm{T}} - (\mathbf{D}^{1/2}\mathbf{V}^{\mathrm{T}})\mathbf{X}^{\mathrm{T}}\mathbf{U} \right\|^2 + \lambda_1 \|\mathbf{U}\|_1 + \lambda_2 \|\mathbf{U}\|^2$$
$$= \min \left\| \mathbf{M}^* - \mathbf{X}^*\mathbf{U}^* \right\|^2 + \lambda \|\mathbf{U}^*\|_1 \tag{9}$$

where $\mathbf{M}^* = \begin{bmatrix} (\mathbf{V}\mathbf{D}^{1/2})^{-1}\mathbf{M}^{\mathrm{T}} \\ \mathbf{0}^{m \times d} \end{bmatrix}$, $\mathbf{U}^* = \sqrt{1+\lambda_2}\,\mathbf{U}$, $\mathbf{X}^* = (1+\lambda_2)^{-1/2} \begin{bmatrix} (\mathbf{D}^{1/2}\mathbf{V}^{\mathrm{T}})\mathbf{X}^{\mathrm{T}} \\ \sqrt{\lambda_2}\mathbf{I}^{m \times m} \end{bmatrix}$, $\lambda = \dfrac{\lambda_1}{1+\lambda_2}$,

$\mathbf{M}^* = [\mathbf{m}_1^*, \mathbf{m}_2^*, \cdots, \mathbf{m}_d^*] \in \mathbf{R}^{(n+m) \times d}$, $\mathbf{U}^* = (\mathbf{u}_1^*, \mathbf{u}_2^*, \cdots, \mathbf{u}_d^*) \in \mathbf{R}^{m \times d}$, and $\mathbf{X}^* = [\mathbf{x}_1^*, \mathbf{x}_2^*, \cdots, \mathbf{x}_m^*] \in \mathbf{R}^{(n+m) \times m}$.

In (9), we want to learn a $d$-dimensional feature matrix $\mathbf{Y}^* = \mathbf{X}^*\mathbf{U}^*$ with sparse restrictions for each projection vector in $\mathbf{U}^*$. We can find the optimal solution of $\mathbf{U}^*$ by finding the optimal value for each $\mathbf{u}_i^*$ for $i = 1, \cdots, d$. Rewrite (9) as

$$\min \left\| \mathbf{M}^* - \mathbf{X}^*\mathbf{U}^* \right\|^2 + \lambda \|\mathbf{U}^*\|_1$$
$$= \min \sum_{i=1}^{d} \left( \left\| \mathbf{m}_i^* - \mathbf{X}^*\mathbf{u}_i^* \right\|^2 + \lambda \|\mathbf{u}_i^*\|_1 \right) \tag{10}$$
$$= \sum_{i=1}^{d} \left( \min \left( \left\| \mathbf{m}_i^* - \mathbf{X}^*\mathbf{u}_i^* \right\|^2 + \lambda \|\mathbf{u}_i^*\|_1 \right) \right).$$

It shows that solving problem (10) equals to find the optimal solution for each of its sub-problem,

$$\min \left( \left\| \mathbf{m}_i^* - \mathbf{X}^*\mathbf{u}_i^* \right\|^2 + \lambda \|\mathbf{u}_i^*\|_1 \right). \tag{11}$$

With problem (11), LARS [Efron et al. 2004] can be applied to find the optimal solution. LARS is an efficient algorithm for solving the lasso penalized multiple linear regression problem. In this paper, we implement LARS carefully to solve STL. Below, we brief LARS for STL. For convenience, we use simple notations (with superscript and subscript removed) for discussions.

To solve the problem

$$\min \left( \|\mathbf{m} - \mathbf{X}\mathbf{u}\|^2 + \lambda \|\mathbf{u}\|_1 \right), \tag{12}$$

LARS iteratively finds the features, which have the largest correlation with the objective function defined in (12) without the 1-norm penalty, and adds them into an active set $\mathcal{A}$. The $\mathbf{y} = \mathbf{X}\mathbf{u}$ is initialized with all items zero and then is iteratively updated according to $\mathcal{A}$. LARS repeats updating $\mathcal{A}$ and $\mathbf{y}$ until $K$ features are included in $\mathcal{A}$, i.e., $K$-sparse ($K$ nonzero entries) $\mathbf{u}$ is achieved.

In particular, LARS begins with $\mathbf{y} = \mathbf{0}$ and then a feature (a column vector in $\mathbf{X}$) which is most correlated with the objective function in (12) without 1-norm penalty is selected and added into the active set $\mathcal{A}$. The correlation $\mathbf{c}$ is defined as the negative gradient of the objective function,

$$\mathbf{c} = -\frac{\partial \left( \|\mathbf{m} - \mathbf{X}\mathbf{u}\|^2 \right)}{\partial \mathbf{u}} = 2\mathbf{X}^{\mathrm{T}}(\mathbf{m} - \mathbf{X}\mathbf{u}) = 2\mathbf{X}^{\mathrm{T}}(\mathbf{m} - \mathbf{y}).$$

The $j = \arg\max |c_j|$ is added to $\mathcal{A}$ and then $\mathbf{y}$ is updated in the direction of $\mathbf{x}_j$ until another feature, whose correlation is equivalent to that of features in $\mathcal{A}$, is added. When there are more than one feature in $\mathcal{A}$, LARS updates $\mathbf{y}$ in a direction equiangular between all features.

We now discuss how to obtain the equiangular vector with the active set $\mathcal{A}$. The matrix $\mathbf{X}_{\mathcal{A}} = [\cdots s_j \mathbf{x}_j \cdots]_{j \in \mathcal{A}}$ consists of all features in $\mathcal{A}$ with $s_j = \text{sign}\{c_j\}$. The $\mathbf{G}_{\mathcal{A}} = \mathbf{X}_{\mathcal{A}}^{\mathrm{T}}\mathbf{X}_{\mathcal{A}}$ is the gram matrix over $\mathbf{X}_{\mathcal{A}}$ and $\mathbf{A}_{\mathcal{A}} = (\mathbf{e}_{|\mathcal{A}|}^{\mathrm{T}}\mathbf{G}_{\mathcal{A}}^{-1}\mathbf{e}_{|\mathcal{A}|})^{-1/2}$, wherein $\mathbf{e}_{|\mathcal{A}|}$ is a vector of 1's with length $|\mathcal{A}|$. The equiangular vector, i.e., the unit vector making equal angles, is estimated by $\mathbf{y}_{\mathcal{A}} = \mathbf{A}_{\mathcal{A}}\mathbf{X}_{\mathcal{A}}\mathbf{G}_{\mathcal{A}}^{-1}\mathbf{1}_{\mathcal{A}}$.

Then, $\mathbf{y}$ is updated as $\mathbf{y}^{new} = \mathbf{y} + \gamma \mathbf{y}_\mathcal{A}$. The $\gamma$ is determined by $\gamma = \min_{j \in \mathcal{A}^c}^+ \left\{ \frac{\tilde{c} - c_j}{A_\mathcal{A} - a_j}, \frac{\tilde{c} + c_j}{A_\mathcal{A} + a_j} \right\}$ where $a = [a_1, \cdots, a_m]^T = \mathbf{X}^T \mathbf{y}_\mathcal{A}$ and $\tilde{c} = \max_j \{|c_j|\}$. The $\mathcal{A}^c$ is the complement of $\mathcal{A}$ and $\min^+$ denotes that the minimum is operated over only positive components within each choice of $j$. With the new $\mathbf{y}$, we can further find the next feature that has the correlation $\tilde{c} = \max_j \{|c_j|\}$ and add it into $\mathcal{A}$. The iterative process is ended until $K$ features are included in $\mathcal{A}$, i.e., $K$-sparse ($K$ nonzero entries) $\mathbf{u}$ is achieved.

4. DISCUSSION – PDA VS. BDA

Section 3.1 introduces PDA to exploit the user feedback knowledge. PDA and BDA share an identical assumption, i.e., all relevant samples are similar to each other and each irrelevant sample is irrelevant in its own way. With this assumption, it is quite straightforward to directly require relevant-relevant sample pairs to be close and relevant-irrelevant pairs to be far away in the projection subspace.

However, instead of pair-wise constrains, BDA requires all relevant samples to be close to the center of the relevant samples and all irrelevant samples to be far away from this center. The objective function of BDA is

$$\arg\max \frac{\mathrm{tr}(\mathbf{U}^T \mathbf{S}_- \mathbf{U})}{\mathrm{tr}(\mathbf{U}^T \mathbf{S}_+ \mathbf{U})}, \tag{13}$$

where the scatter of query irrelevant samples is defined by $\mathbf{S}_- = \sum_{i \in -}(\mathbf{x}_i - \mathbf{m}_+)(\mathbf{x}_i - \mathbf{m}_+)^T$ and the scatter of the query relevant samples is defined by $\mathbf{S}_+ = \sum_{i \in +}(\mathbf{x}_i - \mathbf{m}_+)(\mathbf{x}_i - \mathbf{m}_+)^T$. The mean of relevant samples is $\mathbf{m}_+ = \frac{1}{N_+} \sum_{\mathbf{x}_i \in +} \mathbf{x}_i$.

To find out the relationship between PDA and BDA, we rewrite $\mathrm{tr}(\mathbf{U}^T \mathbf{S}_- \mathbf{U})$ in (13) as

$$\begin{aligned}
\mathrm{tr}(\mathbf{U}^T \mathbf{S}_- \mathbf{U}) &= \mathrm{tr}\left( \mathbf{U}^T \left( \sum_{i \in -}(\mathbf{x}_i - \mathbf{m}_+)(\mathbf{x}_i - \mathbf{m}_+)^T \right) \mathbf{U} \right) \\
&= \sum_{i \in -} \mathrm{tr}\left( \mathbf{U}^T (\mathbf{x}_i - \mathbf{m}_+)(\mathbf{x}_i - \mathbf{m}_+)^T \mathbf{U} \right) \\
&= \sum_{i \in -} \mathrm{tr}\left( (\mathbf{y}_i - \mathbf{m}_+^y)(\mathbf{y}_i - \mathbf{m}_+^y)^T \right) \\
&= \sum_{i \in -} (\mathbf{y}_i - \mathbf{m}_+^y)^T (\mathbf{y}_i - \mathbf{m}_+^y)
\end{aligned} \tag{14}$$

where $\mathbf{m}_+^y = \frac{1}{N_+} \sum_{I_i \in +} \mathbf{y}_i$ is the center of the relevant samples in the learned subspace.

Substitute $\mathbf{m}_+^y$ into (14) and we get

$$\begin{aligned}
\mathrm{tr}(\mathbf{U}^T \mathbf{S}_- \mathbf{U}) &= \frac{1}{(N_+)^2} \sum_{I_i \in -} \left( \left( \sum_{I_m \in +} (\mathbf{y}_i - \mathbf{y}_m) \right)^T \left( \sum_{I_m \in +} (\mathbf{y}_i - \mathbf{y}_m) \right) \right) \\
&= \frac{1}{(N_+)^2} \sum_{I_i \in -} \left( \sum_{I_m \in +, I_n \in +} (\mathbf{y}_i - \mathbf{y}_m)^T (\mathbf{y}_i - \mathbf{y}_n) \right) \\
&= \frac{1}{(N_+)^2} \sum_{I_i \in -} \left( \sum_{I_j \in +} \|\mathbf{y}_i - \mathbf{y}_j\|^2 + \sum_{\substack{I_m \in +, \\ I_n \in +, m \neq n}} (\mathbf{y}_i - \mathbf{y}_m)^T (\mathbf{y}_i - \mathbf{y}_n) \right).
\end{aligned} \tag{15}$$

Similarly, we can rewrite $\text{tr}(\mathbf{U}^T \mathbf{S}_+ \mathbf{U})$ in (13) as

$$\text{tr}(\mathbf{U}^T \mathbf{S}_+ \mathbf{U}) = \frac{1}{(N_+)^2} \sum_{I_i \in +} \left( \sum_{I_j \in +} \|\mathbf{y}_i - \mathbf{y}_j\|^2 + \sum_{\substack{I_m \in +, \\ I_n \in +, m \neq n}} (\mathbf{y}_i - \mathbf{y}_m)^T (\mathbf{y}_i - \mathbf{y}_n) \right). \quad (16)$$

Comparing (15) and (16) against the two parts in (1), i.e., $\frac{1}{N_+ N_-} \sum_{I_i \in +} \sum_{I_k \in -} \|\mathbf{y}_i - \mathbf{y}_k\|^2$ and $\frac{1}{N_+ N_+} \sum_{I_i \in +} \sum_{I_j \in +} \|\mathbf{y}_i - \mathbf{y}_j\|^2$, the difference between PDA and BDA is that PDA has no cross items, i.e., $(\mathbf{y}_i - \mathbf{y}_m)^T (\mathbf{y}_i - \mathbf{y}_n)$.

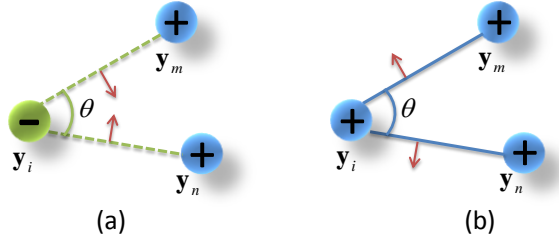

Fig. 3. The cross items in BDA. (a) and (b) correspond to that in (15) and (16) respectively. It shows that the local structure (geometry) cannot be persevered after dimension reduction.

The cross items in BDA intrinsically destroy the local geometry in the projected subspace. As illustrated in Fig. 3, in (15), it maximizes $(\mathbf{y}_i - \mathbf{y}_m)^T (\mathbf{y}_i - \mathbf{y}_n)$ for irrelevance sample $\mathbf{y}_i$ which means that a small angle $\theta$ in Fig.3 (a) is expected. And, in (16), a large angle $\theta$ is required for a relevance sample $\mathbf{y}_i$ by minimizing $(\mathbf{y}_i - \mathbf{y}_m)^T (\mathbf{y}_i - \mathbf{y}_n)$. In recent years, it has been well acknowledged that the local geometry is very useful in dimension reduction and many dimension reduction methods have been developed to preserve the local geometry, e.g., LLE [Roweis and Saul 2000]. Therefore, STL uses PDA to optimize the distance between pairs of samples to retain the local geometry.

5. COMPLEXITY ANALYSIS

For a query, $N$ video shots are returned by the text-based search in which $N_L$ samples are labeled in the feedback process. The complexity consists of four parts,

a) the calculation of $\mathbf{M}$ via PCA decomposition: the time cost is $O(m^2 N + N^3)$ with $O(m^2 N)$ for calculating the covariance matrix and $O(N^3)$ for eigen-decomposition.

b) the calculation of $\mathbf{L}$: it takes $O(N_L^2)$ to derive $\mathbf{L}_i$ for each labeled relevant sample and thus the time cost for $\mathbf{L}$ is $O(N_+ N_L^2)$ where $N_+$ is the number of labeled relevant samples.

c) the calculation of $\mathbf{M}^*$ and $\mathbf{X}^*$: the time cost for computing $\mathbf{M}^*$ and $\mathbf{X}^*$ are $O(N^2 d)$ and $O(N^2 m)$, respectively. Since $d \ll m$, the time cost in this part can be approximated by $O(N^2 m)$.

d) solving $d$ sub-problems defined in (11) via LARS: in most steps of LARS, simple matrix computations are necessary. The size of the active set $\mathcal{A}$ is augmented from 0 to $K$ when $K$-sparse coefficient vector is required. We first analyze the time complexity for the kth round with $|\mathbf{A}^k| = k$ ($k = 1, \cdots, K$) and then sum the time cost over all $K$ rounds. The Gram matrix $\mathbf{G}_{\mathcal{A}^k}$ can be updated incrementally with existing $\mathbf{G}_{\mathcal{A}^{k-1}}$ and then the time cost can be reduced from $O(Nk^2)$ to $O(N(k+1))$. The inverse matrix $\mathbf{G}_{\mathcal{A}^k}^{-1}$ also can be updated from $\mathbf{G}_{\mathcal{A}^{k-1}}^{-1}$ with complexity reduced from $O(k^3)$ to $O(k^2+5k)$. Other operations in LARS, e.g., correlation estimation, take $O((m+1)N)$. We can obtain that, in kth round, the overall time cost is $O(mN+kN+k^2+N)$. By summing over $K$ rounds, the overall time complexity for LARS is $O(mNK+K^3+NK^2)$. Since $K$ is usually much smaller than $m$ and $N$, the time complexity for LARS can be approximately by $O(mNK)$.

Taking all the four parts into account, the total time complexity for STL is $O(m^2N+N^3+N_+N_L^2+mN^2+dmNK)$. Since $N_+$ and $N_L$ are much smaller than $N$, the time cost can be approximated by $O(m^2N+N^3+mN^2+dmNK)$. It is worth emphasizing that the $d$ sub problems in part d) can be processed in parallel for efficiency and then the time cost is reduced to $O(m^2N+N^3+mN^2)$. The time cost for PCA and BDA is $O(m^2N+N^3)$ and $O(m^2N_L+m^3)$ respectively. We can see that, when $m<N$ the time cost of STL is comparable to that of PCA and if $m>N$ the time cost of STL is comparable to that of BDA.

For space cost, we analyze it according to the above 4 parts as well. In a) and b), we need $m \times m$ to store the covariance matrix, $N \times d$ for $\mathbf{M}$ and $N \times N$ for $\mathbf{L}$. We require $N \times d$ and $N \times (m+1)$ to store $\mathbf{M}^*$ and $\mathbf{X}^*$, respectively. In LARS, it needs $K \times K$ for both $\mathbf{G}_{\mathcal{A}}$ and its inverse matrix $\mathbf{G}_{\mathcal{A}}^{-1}$. Therefore, the overall space cost of STL is $O(N^2+mN+m^2)$. Compared with the space cost $O(mN+m^2)$ for both PCA and BDA, STL only need $O(N^2)$ additional space cost. In reranking problem, the space cost of STL is acceptable.

6. EXPERIMENTS

To test the effectiveness of the STL based interactive video search reranking, we conducted extensive experiments on widely used video search benchmark dataset – TRECVID 2005-2007 test set [TRECVID]. This dataset consists of 508 videos and 143,392 shots. The text-based search baseline is obtained based on the Okapi BM-25 formula [Robertson et al. 1997] by using ASR/MT transcripts at shot level. For each of the 72 queries, 24 for each year, the top 1400 shots returned by the search system are used as the initial text-based search result for subsequent reranking. Top 20 samples returned by the text-based search are labeled as relevant or irrelevant samples to mimic the interaction between users and search engines. Figure 4 shows some example video shots and the query relevant ones are marked by " √ ".

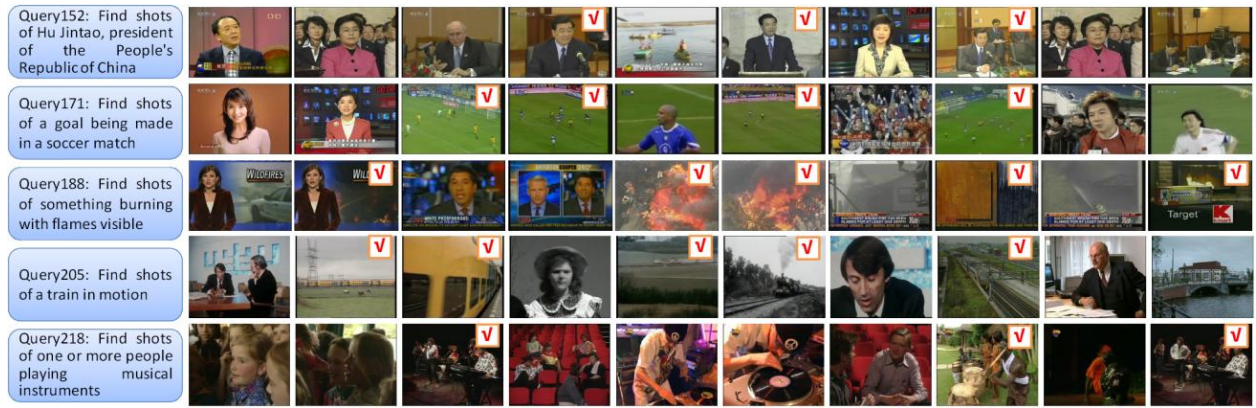

Fig. 4. Example video shots for TRECVID 2005-2007 automatic text search task. For each query, top-10 video shots returned by the text-based search are given and the relevant samples are marked by red " √ ".

In this paper, we utilize four kinds of visual features, i.e., 225-dimensional color moments (CM) [Ma and Zhang 1998], 64-dimensional HSV color histogram (HSV) [Kotoulas and Andreadis 2003], 144-dimensional color auto-correlogram (Corre) [Huang et al. 1997], and 128-dimensional Wavelet texture (WT) [Chang and Kuo 1993]. CM is extracted over 5*5 fixed grid partitions with each block described by a 9-dimensional feature in the Lab color space. HSV color histogram reflects the color distribution characters in the HSV color space. Corre is extracted based on 36 bin color histogram and 4 different distances (1, 3, 5 and 7). These features represent the visual information of the video shots from different aspects and complement with each other. We concatenate them into a long vector with 561 dimensions.

The interactive video search reranking is a general framework. With the visual feature learned by using STL, any reranking method can be adopted by using this query dependent feature. In this paper, we directly use the simplest way to conduct reranking to verify the effectiveness of STL. We rerank the initial text-based search result by calculating the minimal distance of each sample to the labeled relevant samples. For each sample $I_i$ its distance $d_i$ to the query is calculated as

$$d_i = \min_{I_j \in +} D_M(\mathbf{y}_i, \mathbf{y}_j),$$

where $D_M(\mathbf{y}_i, \mathbf{y}_j)$ is the Mahalanobis distance between $\mathbf{y}_i$ and $\mathbf{y}_j$. Then, the reranked result is obtained by ranking sample according to their distances in descending order.

The video shot's relevance is provided by NIST [TRECVID] on two levels, i.e., "Relevant" and "Irrelevant". The most often used performance measure for this dataset is the non-interpolated Average Precision (AP) [Trec measures], which is also adopted in this paper. We average the APs over all 24 queries in each year to get the Mean AP (MAP) for measuring overall performance. For all methods, the parameters are selected via 3-fold cross-validation, i.e., queries in two years are used for validation and the queries in the remaining year are used for test.

6.1 Overall Performance Comparison and Analysis

Table 1 MAP comparison between reranking with STL and other dimension reduction methods on TRECVID 2005

| Method | Text Baseline | SML | SDA | BDA | PCA | STL |
|---|---|---|---|---|---|---|
| MAP | 0.0441 | 0.0762 | 0.0818 | 0.0875 | 0.0912 | **0.0951** |
| Gain | - | 72.79% | 85.49% | 98.41% | 106.80% | **115.65%** |

Table 2 MAP comparison between reranking with STL and other dimension reduction methods on TRECVID 2006.

| Method | Text Baseline | SML | SDA | BDA | PCA | STL |
|---|---|---|---|---|---|---|
| MAP | 0.0381 | 0.0558 | 0.0583 | 0.0554 | 0.0590 | **0.0593** |
| Gain | - | 46.46% | 53.02% | 45.41% | 54.86% | **55.64%** |

Table 3 MAP comparison between reranking with STL and other dimension reduction methods on TRECVID 2007.

| Method | Text Baseline | SML | SDA | BDA | PCA | STL |
|---|---|---|---|---|---|---|
| MAP | 0.0306 | 0.0682 | 0.0710 | 0.0703 | 0.0638 | **0.0734** |
| Gain | - | 122.88% | 132.03% | 129.74% | 108.50% | **139.87%** |

Table 4 The mean performance comparison between reranking with STL and other dimension reduction methods over TRECVID 2005-2007, i.e., the mean performance over three years.

| Method | Text Baseline | SML | SDA | BDA | PCA | STL |
|---|---|---|---|---|---|---|
| MAP | 0.0376 | 0.0667 | 0.0677 | 0.0711 | 0.0713 | **0.0760** |
| Gain | - | 77.39% | 80.05% | 89.10% | 89.63% | **102.13%** |

To test the effectiveness of the proposed STL based interactive video search reranking, we compared it against the text-based search result and existing top dimension reduction methods, including the unsupervised one, e.g., PCA [Hotteling 1933], the supervised one, e.g., BDA [Zhou and Huang 2001], the semi-supervised methods, e.g., SDA [Cai et al. 2007] and SML [Lin et al. 2005]. Tables 1-4 show the experimental results over each year of TRECVID 2005-2007 as well as the mean performance over the three years (denoted by "Over 2005-2007"). The "Text Baseline" denotes the performance of the text-based search result without reranking.

According to Tables 1-3, the proposed STL method outperforms other methods consistently over each of the three years respectively. Table 4 gives the overall performance over 2005-2007 and STL shows significant improvements, i.e., 102.13% related gain over "Text Baseline". It reflects STL can effectively propagate the feedback knowledge from labeled samples to unlabeled ones by using distribution knowledge. For other dimension reduction methods, the unsupervised ones utilize only the sample distribution knowledge and the supervised methods utilize only the user feedback knowledge. Therefore, they cannot perform as well as STL. For the semi-supervised algorithms, i.e., SML and SDA, although they take both distribution and feedback knowledge into consideration, they are developed for the common retrieval problem and don't take the characteristics of the reranking problem into consideration. Thus, they cannot generalize well for the video search reranking. In addition, sparsity can make the projection matrix more succinct and simpler, can reduce the over-fitting problem, and can provide a good interpretation of the projection matrix. Besides, in STL, the sparse projection matrix reduces the computation and storage cost for the subsequent process. Therefore, we prefer STL in interactive video search reranking.

6.2  Performance Comparison over Each Query

Section 6.1 shows the overall performance in terms of MAP over each year. Besides the overall performance, we also investigated the performance over each query for STL and compare them with Text Baseline, SML, SDA, BDA and PCA. The results are presented in Fig. 5. We can see that reranking with all dimension reduction tools can be improved significantly compared against the text-search baseline on most queries. It verifies the effectiveness of the STL based interactive video search reranking framework.

It is well-known that TRECVID dataset is challenging and there are often rare relevant samples for the query. The poor performance of the text-based search baseline confirms this point. By investigating the numbers of relevant samples returned in the text-based search, we find that BDA usually performs well on those queries which have good text-based search baselines. This is because a better text-based search baseline returns more relevant samples in the top 20 ranked samples. In other words, we can get more labeled relevant samples for BDA to learn the discriminative information. On the other hand, PCA achieves good performance on those queries which have sufficient relevant samples. This is because PCA obtains more reliable knowledge by learning the salient information contained in the dataset, when the number of relevant samples is large.

BDA and PCA show their strengths on certain queries because they leverage the knowledge of labeled and unlabeled data respectively. STL takes merits of both labeled and unlabeled knowledge and thus achieves much better performance than them. Figure 5 shows that STL performs better than or at least comparable to BDA and PCA steadily on most queries. For those queries, in which one of BDA and PCA performs well and the other performs bad such as query178 ("*Find shots of US Vice President Dick Cheney*"), STL can achieve a moderate performance by learning a good balance between the two items. Therefore, STL learns the knowledge in both labeled and unlabeled domain, taking the advantages of both, and gives a good performance. For semi-supervised method SML and SDA, although they use both the labeled and unlabeled data, as discussed in Section 2, they are not designed specifically for video search reranking problem and improper manifold assumption in these methods leads to limited performance improvement in reranking.

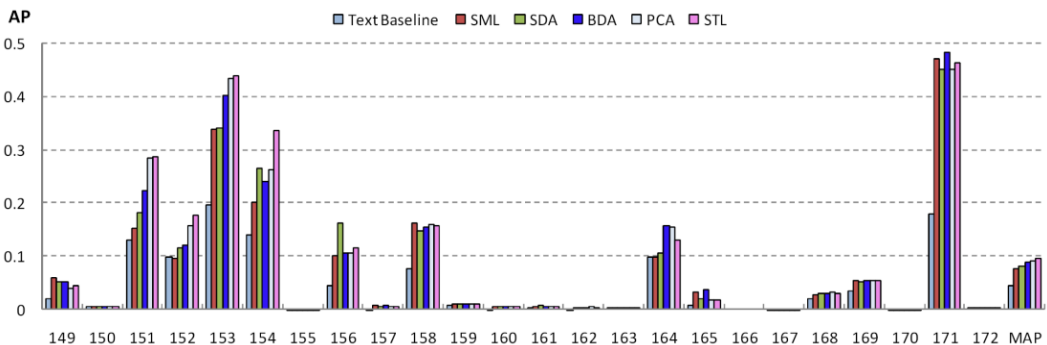

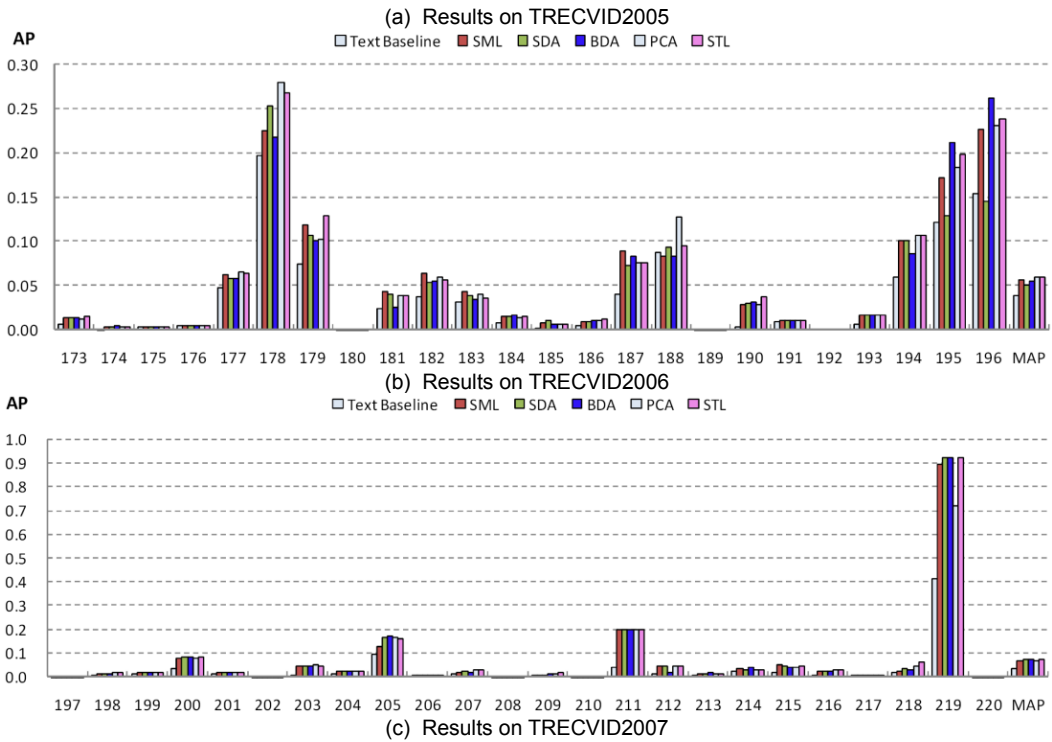

Fig. 5. Experimental results in terms of AP over each query on TRECVID 2005-2007. On TRECVID2005 and TRECVID2007, STL outperform the Text Baseline as well as other methods on most of the queries.

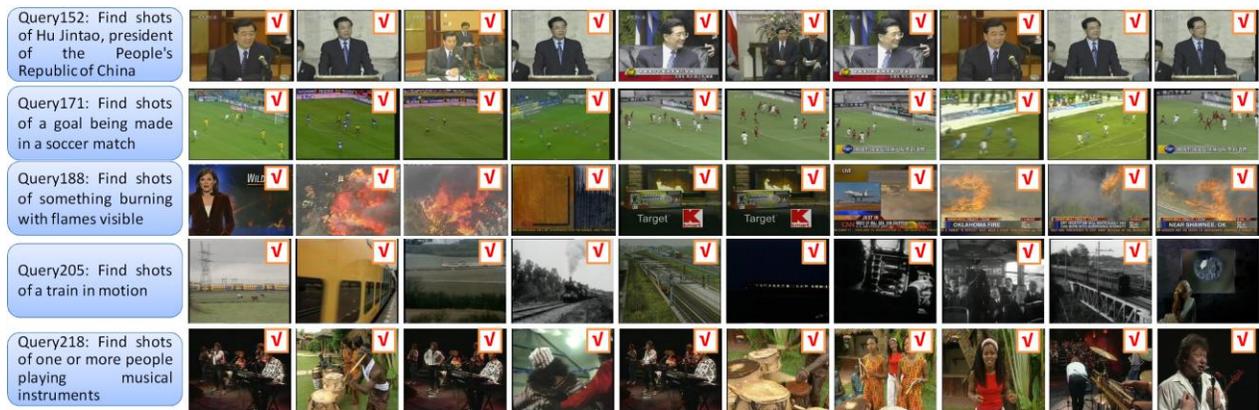

Fig. 6. Reranking results of the five queries listed in Fig. 4. For each query, the top-10 video shots returned by STL are given with relevant samples marked by red "√".

Figure 6 gives reranking results of the five queries in Fig. 4. For each query, the top-10 samples returned by SML are presented with relevant samples marked by red "√". Compared with the text-based search result given in Fig. 4, we can see that the performances are vastly improved by using SML.

For STL, we further show the coefficient path of the LARS in the sparse learning process by taking query211 and query212 for illustration. Figure 7 shows the coefficient path of learning the 10th basis of the projected features, i.e., the 10th column of the projection matrix. Several relevant and irrelevant video shots for these two queries are also given. In LARS, all entries of the column vector are zeros initially and then 561 features from four categories (i.e., CM, HSV, Corre and WT), where each category is indicated by a mark, are sequentially added into the active set according to their importance. The more important features are selected firstly and be assigned larger value than those which are less important. Figure 7(a) shows CM and Corre features are more effective than HSV and WT for "*sheep*" and "*goats*". And, Fig. 7(b) shows that WT is effective for "*boat moves past*" while CM has less impact on this query. By viewing the relevant and irrelevant video shots given in Fig. 7(a) and (b), we can see that in query211 the "*sheep*" and "*goats*" are usually in white color, therefore the color related feature CM and Corre are more effective. For query212, the relevant video shots with "*boat*" usually also contain a lot of water. The water has good textual pattern, as a consequence, textual related feature WT plays more important role. From these, we can see that the feature selection process in sparse learning can help us to find the contextual information, can help us to understand the query better, and can give a guideline for developing more effective visual features for reranking.

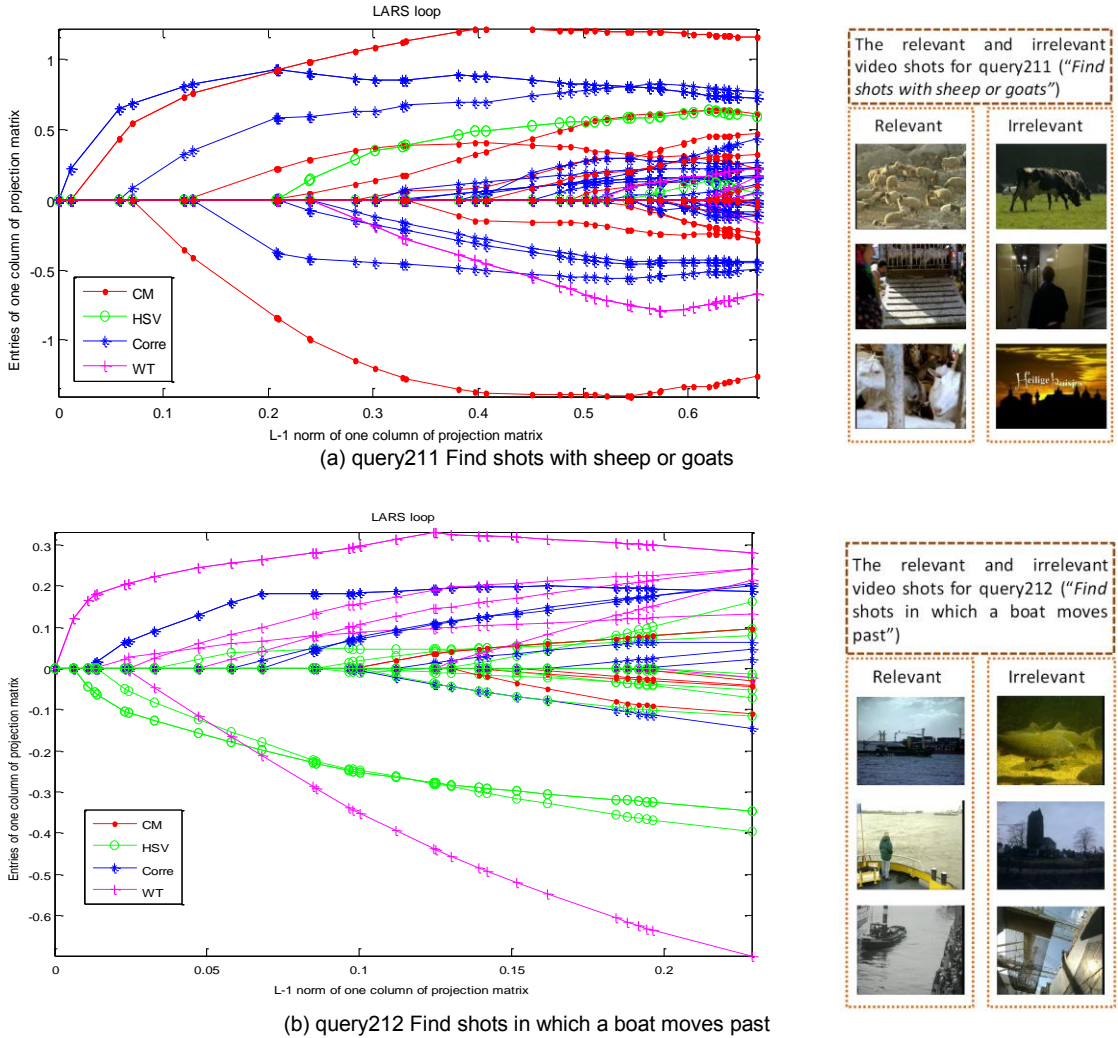

(a) query211 Find shots with sheep or goats

(b) query212 Find shots in which a boat moves past

Fig. 7. Coefficient path of LARS for STL, *i.e.*, the entries of the column of projection matrix *vs.* its 1-norm in LARS, in query211. The coefficient path shows that CM and Corre are more important than HSV and WT in encoding the visual information to represent the user's intention.

## 7. CONCLUSION

In this paper, interactive video search reranking is utilized to incorporate user's labeling information for significantly improving the effectiveness of the visual search reranking. User's intention, represented by labeled samples, can be encoded in a newly proposed sparse transfer learning (STL). STL efficiently and effectively exploits the user feedback knowledge and the sample distribution knowledge from different domains and then transfer user's intention from labeled samples to unlabeled samples. The sparsity property is incorporated to derive a more compact representation in the learned subspace. We have conducted extensive experiments on the benchmark video search dataset TRECVID 2005 – 2007. The experimental results demonstrate the effectiveness of the proposed method. In the future, we will apply the learned coefficient path for result representation to benefit user interface design.